\documentclass{andp2012}
\usepackage[english]{babel}
\usepackage{bm,graphicx}


\title{Quantum Pairing Time Orders}

\author[F.\,A.V. Balatsky]{A.V. Balatsky\inst{1,2}\footnote{Corresponding author\quad E-mail:~\textsf{avb@nordita.org}}}
\author[S.\,P.O. Sukhachov]{P.O. Sukhachov\inst{1}}
\author[T.\,S. Bandyopadhyay]{S. Bandyopadhyay\inst{1}}
\address[1]{
Nordita, KTH Royal Institute of Technology and Stockholm University, Roslagstullsbacken 23, SE-106 91 Stockholm, Sweden
}
\address[2]{Department of Physics, University of Connecticut, Storrs, CT 06269, USA}

\shortauthors{A.V. Balatsky, P.O. Sukhachov, S. Bandyopadhyay}

\begin{abstract}
We propose the concept of the time-independent correlators for the even- and odd-frequency pairing states that can be defined for both bosonic and fermionic quasiparticles. These correlators explicitly capture the existence of two distinct classes of pairing states and provide a direct probe of the hidden Berezinskii order.
This concept is illustrated in the cases of pairings for Majorana fermions and quasiparticles in Dirac semimetals.
It is shown that the time-independent correlator is able to effectively capture the energy scale relevant for pairing.
\end{abstract}
\shortabstract

\begin{document}
\maketitle

\section{Introduction}
\label{sec:intro}

Questions on the nature of order of quantum matter in time domain become an exciting frontier in quantum condensed matter. Introduction of time dynamics increases the complexity of the problem of quantum correlations and orders. The very concepts of equilibrium physics are no longer applicable. For example, the singular role the ground state plays in determining the properties system is not relevant in dynamical systems \cite{Rudner2013,Bukov2015}. Important questions being raised about the meaning of the phase transitions and  new phases that would emerge as a result of active drive of quantum states. One of the prominent examples is the discovery of time crystals, initially proposed by F. Wilczek \cite{Wilczek2012TX,TXreview}.

Questions of classification of quantum time orders is an important challenge that community addresses now. One possibility in this conversation about quantum coherences developed in time is to look at the correlations that are induced by explicitly driving the system. Alternative to the external drive would be nontrivial time correlations that develop in materials. The purpose of this article is to address the following question: is it possible to have an inherently dynamic phase in equilibrium? We now know that the answer to the question on existence of nontrivial time order in equilibrium phases is positive. The phase that  realizes nontrivial temporal order in equilibrium is the odd-frequency pairing state \cite{Tanaka-Nagaosa:rev-2012,LinderRMP,Triola-Black-Schaffer:rev-2019}. One thus can conjecture that odd-frequency order is connected to the dynamic orders discussed in the context of driven quantum dynamics.

At first, the question seem paradoxical. How one can have a dynamic phase in equilibrium?  What is meant is to seek the orders that have nontrivial dependence on relative times of the two-particle correlation functions that used to describe the order. To make the point, we focus on the class of superconducting states that exhibit pairing correlations that develop a node in relative  time, the so-called odd-frequency states. These superconducting states were posited by Berezinskii in 1974 in the context of fermion pairing in the superfluid He$^3$~\cite{Berezinskii:1974}, for a review, see ref.~\cite{LinderRMP}.

Consider orders that can be described by broken symmetries with respective correlations describable by two-particle coherences. Specifically we look at the pairing state described by correlations functions of two fermions $F(\bf{r}_1,\tau_1|\bf{r}_2, \tau_2) = \langle \psi_{\bf{r}_1}( \tau_1)\psi_{\bf{r}_2}(\tau_2)\rangle $. Reasons one prefers to call these state pairing states instead of superconducting state is the fact that these pairing states can develop for neutral particles like superfluid pairing and Majorana fermion pairing correlations where one has pairing but there is no flow of any current nor other attributes of a charged superconductor.

Once the state develops the particle-particle condensate, we focus on the classification of the pairing states with respect to a point group of crystal $\mathbb{G}$ and translations $\mathbb{T}$. An ordered state breaks the initial group to an invariant subgroup that describes the correlated state. One can then classify the states as conventional or unconventional pairing states depending on the symmetry properties and what symmetry was broken. For example, if the pairing correlator breaks the initial point group, we find the unconventional pairing states sometimes called {\it p-wave} or {\it d-wave} state, depending on the symmetry of the order \cite{SigristRMP}. Similarly, the translational symmetry can be broken in the paired state and one can claim the translationally noninvariant states like the Fulde–Ferrell–Larkin–Ovchinnikov (FFLO) state and pair density waves \cite{FFLO}. The majority of the analysis of pairing states was done in that mode.

One can analyze the time-dependent orders in a similar fashion and seek the classification of pairing states with respect to time. Turns out that there are two suspect possibilities here:  one is the correlations that are even in relative time $\tau - \tau^{\prime}$ and surprisingly there are pairing states that are odd in $\tau - \tau^{\prime}$. It was Berezinskii~\cite{Berezinskii:1974} who pointed out that a new class of pairing states would emerge if one considers the breakdown of the ``parity" with respect to permutation of time coordinates on pairing correlator. To be precise, we start with the general definition of  pairing correlator
\begin{equation}\label{Eq1}
  F_{\alpha \beta}(\tau, \tau^{\prime}) = \langle T_{\tau} \psi_\alpha(\tau) \psi_\beta(\tau^{\prime})\rangle = F_{\alpha \beta}(\tau - \tau^{\prime}),
\end{equation}
where $0 \geq \tau \geq 1$ is the Matsubara time (we assume that temperature $T=1$), $\psi_\alpha(\tau)$ is the Fermi annihilation operator with index $\alpha$ labeling space, band, spin index. The detailed dependence on these indices is not important as we focus on time dynamics, $T_\tau$ is the time ordering. For equilibrium system, we assume here,  $F_{\alpha \beta}$ depends only on $\tau - \tau^{\prime}$ and we will keep the time dependence on the first operator $F_{\alpha \beta} (\tau) = \langle T_{\tau} \psi_\alpha(\tau) \psi_\beta(0)\rangle$.

The symmetry of pairing states is constrained by the condition that correlator should be odd under simultaneous interchange of all indices:
\begin{equation}
\label{Eq:SPOT1}
  F_{\alpha \beta}(\tau - \tau^{\prime}) = (-1) F_{\beta \alpha}(\tau^{\prime}-\tau).
\end{equation}
This constraint can be rewritten as
\begin{equation}\label{Eq:SPOT2}
  SP^*OT^* = -1,
\end{equation}
where $S$ is the operation  of permutation of spin indices, $P^*$ is the permutation of the coordinates of particles $\bf{r}_1 \leftrightarrow \bf{r}_2$, $O$ is the permutation of orbital indices, and $T^*$ is the permutation of the time coordinates $\tau_1 \leftrightarrow \tau_2$, see e.g. \cite{LinderRMP}. Note that $T^*$ and $P^*$ are starred to indicate that the operations are the permutations and not space and time inversion. The origin of the constraint is in the definition of the time-ordered function for fermions \cite{LinderRMP}.

From the $SPOT$ classification one can infer the existence of the odd frequency (Berezinskii) pairing where $T^* = -1$ and one thus can develop spin-singlet pairing correlations that have a p-wave structure as a function of relative coordinate $S = -$, $P^* = -$, $T^* = -$, $O = +$, which can be shorthand  written as $--+-$ correlations. Alternatively, one can have s-wave triplet pairing correlations, initially proposed by Berezinskii, that have a $+++-$ signature.

An odd-frequency state can be viewed as a hidden order in a sense that the pairing correlations vanish at equal time as a pairing correlation is odd under $\tau \leftrightarrow \tau^{\prime}$ and hence equal time correlation is zero in the Matsubara time. The long-standing question in the field is what is the natural order parameter for an odd-frequency superconductor. One approach, which is often taken in the field, would be to take the time derivative of the pairing correlation function at $\tau \rightarrow 0$.
This paper will present alternative {\it time-independent} correlators that will explicitly capture the existence of two distinct classes of superconductors, see Table~\ref{tab:Tab1}. These two correlation functions, namely $A_{\alpha \beta}$ for even- and $B_{\alpha \beta}$ for the odd-frequency pairing state, respectively, are derived here. This concept can be also generalized to the case of bosons, where the correlators $\mathcal{A}_{\alpha \beta}$ and $\mathcal{B}_{\alpha \beta}$ should be used.

The plan of the paper is to introduce the pairing correlators in Section~\ref{sec:correlators}. In Sections~\ref{sec:Ber-Majorana} and \ref{sec:Ber-Dirac}, we will give  two examples of explicit calculations for these  correlators. First, we consider the case of Majorana fermions that realize the odd-frequency pairing. Second, we discuss the case of the odd-frequency pairing in Dirac semimetals and present $B_{\alpha \beta}$ correlator for this case as well.

\section{Time-independent correlations in even-frequency and Berezinskii channels}
\label{sec:correlators}

We prove the existence of two classes of superconductors that have the same spin quantum numbers, total spin $S = 0,1$ but with the opposite orbital parities. While the key attention will paid to the case of fermions, bosonic systems will be also considered.

\subsection{Fermionic case}
\label{sec:correlators-fermions}

Let us start with a general relation for fermions $F_{\alpha \beta}(\tau) = -F_{\alpha \beta}(\tau +1)$. This results in the Matsubara frequencies that are odd: $\omega_n = (2n+1)\pi$. Hence, any correlator can be represented through Matsubara components.

Let us introduce two types of time-independent correlators:
\begin{eqnarray} \label{Eq:Def1}
  A_{\alpha \beta} &=& \int\limits_0^1 d\tau  \langle T_{\tau} \frac{d}{d\tau}\psi_\alpha(\tau) \psi_\beta(0)\rangle,\\
  \label{Eq:Def1-b}
  B_{\alpha \beta} &=& \int\limits_0^1 d\tau \langle T_{\tau}\psi_\alpha(\tau) \psi_\beta(0)\rangle.
\end{eqnarray}

One can prove that $A_{\alpha \beta}$ is antisymmetric upon permutation of indices and $B_{\alpha \beta}$ is symmetric. We will illustrate the proof for  the correlator $A_{\alpha \beta}$. Starting  with
\begin{eqnarray} \label{Eq.Deriv1}
 A_{\alpha \beta} &=& \int\limits_0^1 d\tau  \langle T_{\tau} \frac{d}{d\tau}\psi_{\alpha}(\tau) \psi_{\beta}(0)\rangle = \nonumber\\
&+&\int\limits_0^1 d\tau\langle \psi_{\alpha}(\tau)[H, \psi_{\beta}](0)\rangle  = \nonumber \\
&-&\int\limits_0^{-1} d\eta \langle \frac{d}{d\eta} \psi_{\beta}(\eta) \psi_{\alpha}(0)\rangle = -A_{\beta \alpha},
\end{eqnarray}
where $H$ is the Hamiltonian, $\eta = - \tau$, and the properties of time-ordered correlation were used. Proof of $B_{\alpha \beta}=B_{\beta \alpha}$ can be performed in a similar fashion.

Thus, the correlators $A_{\alpha \beta}$ and $B_{\alpha \beta}$ defined in Eqs.~(\ref{Eq:Def1}) and ~(\ref{Eq:Def1-b}), respectively, have the following properties:
\begin{eqnarray}
A(s_1, \bf{r}_1 | s_2, \bf{r}_2) &=& - A(s_2, \bf{r}_2 | s_1, \bf{r}_1),  \\
B(s_1, \bf{r}_1 | s_2, \bf{r}_2) &=& B(s_2, \bf{r}_2 | s_1, \bf{r}_1).
\label{Eq.Explic1}
\end{eqnarray}
For a given spin-projected state we have:

a) $S = 0$, spin-singlet state. It follows that there are two families of spin-singlet superconductors, one is with $S = -$, $P^* = -$ for $A_{\alpha \beta}$ correlator and one is for $S = -$, $P^* = +$ for $B_{\alpha \beta}$ correlator. The latter corresponds to the even-frequency superconductor $-+++$  and the other to the odd frequency state with $--+-$ state (in the $SPOT$ nomenclature, the single band case considered here means $O= +$).

b) $S = 1$, spin-triple state. In that case, $B_{\alpha \beta}$ would capture the $+++-$ state and $A_{\alpha \beta} \neq 0$ corresponds to the even-frequency case $+-++$.

This construction thus points to the existence of two types of superconducting states that have corresponding {\em time-independent} correlators $A_{\alpha \beta}$ and $B_{\alpha \beta}$ with opposite parities for same spin states.

It is useful to  introduce:
\begin{eqnarray} \label{Eq:Ap1}
   F^e_{\alpha \beta }(\tau) &=& \sum_{n=0}^\infty F_{\alpha \beta, n}^{e} \cos(\omega_n \tau), \\
   F^o_{\alpha \beta}(\tau) &=& \sum_{n=0}^\infty F_{\alpha \beta, n}^{o} \sin(\omega_n \tau).
\end{eqnarray}
Here $cos(\omega_n \tau)$ and $\sin(\omega_n \tau)$ project even and odd in $\tau$ components from the total $F_{\alpha \beta }(\tau) = F_{\alpha \beta }^e(\tau) + F_{\alpha \beta }^o(\tau)$. The similar relation holds for the Fourier transforms $F_{\alpha \beta, n} = F^{e}_{\alpha \beta, n} + F^{o}_{\alpha \beta, n}$.

Even and odd components have the following properties:
\begin{align}\label{Eq:Ap2}
B_{\alpha \beta} &= 0 \quad &\mbox{for}\quad F^e_{\alpha \beta} \\
B_{\alpha \beta} &= 2 \sum_{n = 0}^\infty \frac{1}{\omega_n} F^o_{\alpha \beta, n} \quad &\mbox{for}\quad  F^o_{\alpha \beta}
 \end{align}
and explicit expressions for $A_{\alpha \beta}$ are:
\begin{align}
\label{Eq:Ap3}
A_{\alpha \beta} &= -2 \sum_{n = 0}^\infty F^e_{\alpha \beta, n} \quad &\mbox{for}\quad F^e_{\alpha \beta} \\
A_{\alpha \beta} &= 0 \quad &\mbox{for}\quad  F^o_{\alpha \beta}
\end{align}

It is worth noting that one can introduce a more general definition of the time-independent correlator for fermions
\begin{equation}
\label{Eq:Def-A-N}
C_{\alpha \beta}^{(N)} = \int\limits_0^1 d\tau  \langle T_{\tau} \frac{d^{N}}{d\tau^{N}}\psi_\alpha(\tau) \psi_\beta(0)\rangle.
\end{equation}
It is related to the correlators $A_{\alpha \beta}$ and $B_{\alpha \beta}$ as $C_{\alpha \beta}^{(0)}\equiv B_{\alpha \beta}$ and $C_{\alpha \beta}^{(1)}\equiv A_{\alpha \beta}$. The correlator given in Eq.~(\ref{Eq:Def-A-N}) has the following even and odd components:
\begin{align}
\label{Eq:Def-A-N-even}
C_{\alpha \beta}^{(N)} &= 2i \sum_{n = 0}^\infty (i\omega_n)^{N-1} F^o_{\alpha \beta, n} \quad &\mbox{for}\quad \mbox{even}\,\, N \\
\label{Eq:Def-A-N-odd}
C_{\alpha \beta}^{(N)} &= -2 \sum_{n = 0}^\infty (i\omega_n)^{N-1} F^e_{\alpha \beta, n} \quad &\mbox{for}\quad  \mbox{odd}\,\, N.
\end{align}
The generalized function (\ref{Eq:Def-A-N}) can be useful to identify the even- and odd-frequency pairings that are not captured by the correlators $B_{\alpha \beta}$ and $A_{\alpha \beta}$.

\begin{table*}[ht]
  \centering
 \begin{tabular}{|c|c|c|c|c|}
  \hline
   Correlator & Fermion even-f & Fermion odd-f & Boson even-f & Boson odd-f \\
  \hline
  $A_{\alpha \beta}$  & $2\sum_{n = 0}^{\infty} F^e_{\alpha \beta, n}$ & 0 & 0 & 0 \\
  \hline
  $B_{\alpha \beta} $ & 0 &  $2\sum_{n = 0}^{\infty} F^o_{\alpha \beta, n}/\omega_n$ & 0 & 0  \\
  \hline
   $\mathcal{A}_{\alpha \beta}$ & 0 & $\sum_{n = 0}^{\infty} \omega_n F^o_{\alpha \beta, n}$ & 0 & $-\sum_{n = 0}^{\infty} \omega_n \mathcal{F}^o_{\alpha \beta, n}$\\\hline
   $\mathcal{B}_{\alpha \beta}$ & $-2\sum_{n = 0}^{\infty} F^e_{\alpha \beta, n}$ & $2\sum_{n = 0}^{\infty} F^o_{\alpha \beta, n}/\omega_n$ & -$\sum_{n = 0}^{\infty} \mathcal{F}^e_{\alpha \beta, n}$ & 0\\
  \hline
\end{tabular}
\caption{Even- and odd-frequency components of the correlation functions $A_{\alpha \beta}$, $B_{\alpha \beta}$, $\mathcal{A}_{\alpha \beta}$, and $\mathcal{B}_{\alpha \beta}$ are presented for a general case of the correlation functions for fermionic $F_{\alpha \beta}$ and bosonic $\mathcal{F}_{\alpha \beta}$ systems. The correlators $A_{\alpha \beta}$ and $B_{\alpha \beta}$ are selective to even and odd pairing channel for fermions but are insensitive to the anomalous pairing channel in bosonic system. On the other hand, the order parameters $\mathcal{A}_{\alpha \beta}$ and $\mathcal{B}_{\alpha \beta}$ can be used to probe bosonic systems.}
\label{tab:Tab1}
\end{table*}

\subsection{Bosonic case}
\label{sec:correlators-bosons}

In this subsection, we generalize the concept of the time-independent correlators $A_{\alpha\beta}$ and $B_{\alpha\beta}$ to the case of bosons. The corresponding correlator $\mathcal{F}_{\alpha\beta}(\tau-\tau^{\prime})=\langle T_{\tau} \phi_\alpha(\tau) \phi_\beta(\tau^{\prime})\rangle$ is even under simultaneous interchange of all indices
\begin{equation}
\mathcal{F}_{\beta\alpha}(\tau^{\prime}-\tau)=(+1)\mathcal{F}_{\alpha\beta}(\tau-\tau^{\prime}).
\end{equation}
As is well known, the Berezinskii classification exists also for bosonic system~\cite{Balatsky:2014-bosons,LinderRMP}. In such a case, the $SPOT$ classification rule becomes identity,
\begin{equation}\label{Eq:SPOT2-boson}
SP^*OT^* = 1,
\end{equation}
from which the possibility of the odd-frequency anomalous pairing with $T^*=-1$ immediately follows.
Such an order parameter could be observed in nematic Bose--Einstein condensate (BEC) systems with multiband or multiorbital hybridization \cite{Balatsky:2014-bosons,LinderRMP}. Unfortunately, the correlators $A_{\alpha\beta}$ and $B_{\alpha\beta}$ used for the case of fermions are trivial for bosons. Indeed, by using the fact that the Matsubara frequency runs over even integers $\omega_n=2n\pi$ and $\mathcal{F}_{\alpha\beta}(\tau)=\mathcal{F}_{\alpha\beta}(\tau+1)$, one can straightforwardly calculate
\begin{eqnarray}
A_{\alpha \beta} &=& \int\limits_0^1 d\tau
\langle T_{\tau} \frac{d}{d\tau} \phi_\alpha(\tau) \phi_\beta(0)\rangle = \ 0,\\
B_{\alpha \beta} &=& \int\limits_0^1 d\tau \langle T_{\tau} \phi_\alpha(\tau) \phi_\beta(0)\rangle = \  0  \quad \mbox{for} \quad n \neq 0.
\end{eqnarray}
Here $\phi_\alpha(\tau)$ is the bosonic field.
For $n=0$, where there is a condensed phase that that corresponds to a zero mode, one
would get a finite
$B_{\alpha\beta}=F_{\alpha \beta}(\tau=0)$. It, however, corresponds only to the even-frequency case. Thus, neither $A_{\alpha\beta}$ nor $B_{\alpha\beta}$ are sensitive to the odd-frequency pairing in bosonic systems.

An efficient way to probe the Berezinskii state is to use the following time-independent order parameters:
\begin{eqnarray}
\mathcal{A}_{\alpha\beta}&=&\lim_{T\rightarrow 1}\int\limits_{0}^{1} d\tau\langle T_{\tau}\frac{d}{d T^{-1}}\frac{d}{d\tau}\phi_{\alpha}(\tau)\phi_{\beta}(0)\rangle,\\
\mathcal{B}_{\alpha\beta}&=&\lim_{T\rightarrow 1}\int\limits_{0}^{1} d\tau\langle T_{\tau}\frac{d}{dT^{-1}}\phi_{\alpha}(\tau)\phi_{\beta}(0)\rangle.
\end{eqnarray}
Physically, they could be viewed as analogs of the heat capacity.
With the above definitions for bosonic systems, we get,
\begin{align}
\mathcal{A}_{\alpha \beta} &= 0 \quad &\mbox{for}\quad  \mathcal{F}^{e}_{\alpha \beta}\\
\mathcal{A}_{\alpha \beta} &=- \sum_{n = 0}^\infty\omega_n \mathcal{F}^{o}_{\alpha \beta, n} \quad &\mbox{for}\quad \mathcal{F}^{o}_{\alpha \beta}
\end{align}
and
\begin{align}
\mathcal{B}_{\alpha \beta} &= - \sum_{n = 0}^\infty \mathcal{F}^{e}_{\alpha \beta, n} \quad &\mbox{for}\quad \mathcal{F}^{e}_{\alpha \beta} \\
\mathcal{B}_{\alpha \beta} &= 0 \quad &\mbox{for}\quad \mathcal{F}^{o}_{\alpha \beta}
\end{align}
Here
\begin{equation}
\mathcal{F}_{\alpha\beta}(\tau)= \sum_{n=0}^{\infty} \mathcal{F}_{\alpha\beta, n}^{e}\cos{\left(\omega_n\tau\right)} + \sum_{n=0}^{\infty} \mathcal{F}_{\alpha\beta, n}^{o}\sin{\left(\omega_n\tau\right)}.
\end{equation}

Like in the case of fermions (see Eq.~(\ref{Eq:Def-A-N})), one can generalize the time-independent bosonic correlators $\mathcal{A}_{\alpha \beta}$ and $\mathcal{B}_{\alpha \beta}$ to the case of arbitrary derivatives over $\tau$:
\begin{equation}
\label{Eq:Def-C-N-bosons}
\mathcal{C}_{\alpha \beta}^{(N)} = \lim_{T\rightarrow 1}\int\limits_0^{1} d\tau  \langle T_{\tau} \frac{d}{d T^{-1}} \frac{d^{N}}{d\tau^{N}}\phi_\alpha(\tau) \phi_\beta(0)\rangle.
\end{equation}
The correlator $\mathcal{C}_{\alpha \beta}^{(N)}$ has the following even and odd components:
\begin{align}
\label{Eq:Def-C-N-bosons-even}
\mathcal{C}_{\alpha \beta}^{(N)} &= (-1)^{\frac{N}{2}+1}\sum_{n = 0}^\infty \omega_n^{N} \mathcal{F}^e_{\alpha \beta, n} \quad &\mbox{for}\quad \mbox{even}\,\, N \\
\label{Eq:Def-C-N-bosons-odd}
\mathcal{C}_{\alpha \beta}^{(N)} &= (-1)^{\frac{N+1}{2}}\sum_{n = 0}^\infty \omega_n^{N} \mathcal{F}^o_{\alpha \beta, n} \quad &\mbox{for}\quad  \mbox{odd}\,\, N.
\end{align}

\section{Berezinskii pairing of Majorana fermions}
\label{sec:Ber-Majorana}

Let us apply the concept of the time-independent correlator $B_{\alpha \beta}$, which can be viewed as an order parameter for odd-frequency pairing, to the specific example of Majorana fermions. Free theory with zero energy Majorana fermions is described by the following Lagrangian: 
\begin{eqnarray}
\label{Eq:Maj1}
L = \mu^\dag \partial_{\tau}\mu.
\end{eqnarray}
In this case, the Berezinskii pairing is natural. Indeed, Majorana fermions are self-adjoint and the creation and annihilation operators are identical $\mu^\dag(\tau)= \mu_(\tau)$.
Then, the single particle propagator $G(\tau) = \langle T_{\tau} \mu^\dag(\tau)\mu(0)\rangle$ for Majorana fermions is identical to anomalous Greens function (pairing correlator), $G(\tau) \equiv F(\tau) = \langle T_{\tau} \mu(\tau) \mu(o)\rangle$.
In terms of Matsubara frequencies,
$G(\omega_n) = F(\omega_n) = 1/(i\omega_n)$ and $F(\tau) = sign(\tau)/2$ for small $\tau$.
Thus the zero energy mode does realize the Berezinskii pairing state (for more discussion, see ref.~\cite{LinderRMP} and references therein). Using Table~\ref{tab:Tab1}, the proper time-independent correlator for the Majorana odd-frequency pairing on the same wire $\alpha=\beta=1$ is defined by
\begin{eqnarray}
B_{11}= 2 \sum_{n=0}^\infty \frac{1}{i\omega_n^2} = -\frac{i}{4}
\label{Eq.Maj2}
\end{eqnarray}
and $A_{11} = 0$.

For the case of interacting Majorana fermions, odd-frequency correlations depend on the nature of interactions. Let us consider the following two cases:

a) Interacting Majorana fermions at the ends of wire. In this case, one would expect that the free Majorana propagator $F(\omega_n) \sim 1/(i\omega_n) $ will be replaced by \begin{equation}
F(\omega_n) \sim \frac{i\omega_n}{\omega_n^2 + J^2},
\label{Eq.Maj2a}
\end{equation}
where $J$ is the interaction scale \cite{Huang2016,LinderRMP}, which could correspond for the intra-wire hybridization. The order parameter $B$ takes form
\begin{eqnarray}
B = i\sum_{n = 0}^\infty \frac{2}{\omega_n^2 +J^2} = \frac{i}{2J} \tanh{\frac{J}{2}}
\label{Eq.Maj4}
\end{eqnarray}
and diminishes as $\sim 1/J$ at large $J$.

b) Sachdev--Ye--Kitaev (SYK) model. This is the model of ``all with all" interactions between many zero-dimensional Majorana fermions~\cite{SYK:1993,SYK:2015,SYK1}. Its Lagrangian reads as
\begin{eqnarray}
\label{Eq.Maj3}
L = \frac{1}{2} \sum_{i} \mu_i \partial_{\tau} \mu_i -\frac{J_{ijkl}}{4!} \sum_{ijkl} \mu_i \mu_j \mu_k \mu_l
\end{eqnarray}
with $\langle J_{ijkl}J_{ijkl} \rangle = 6J^2/N^3$. The large-N limit propagator is~\cite{SYK1}
\begin{eqnarray}
\label{Eq.Maj5}
F(\tau_1, \tau_2) = b\frac{\mbox{sign}(\tau_1- \tau_2)}{\left(J|\tau_1 - \tau_2|\right)^{1/2}},
\end{eqnarray}
where the term with $\partial_{\tau}$ in Eq.~(\ref{Eq.Maj3}) was neglected and $b^4 = 1/4\pi$. The order parameter $B$ reads as
\begin{eqnarray}
\label{Eq.Maj6}
B \simeq -\frac{2b}{J^{1/2}}.
\end{eqnarray}
Is one can see from Eqs.~(\ref{Eq.Maj4}) and (\ref{Eq.Maj6}), the time-independent correlator shows a different scaling with $J$ for different types of interaction between Majorana fermions. Note also that the possibility of the odd-frequency superconductivity in the SYK model was recently considered in refs.~\cite{LinderRMP,Gnezdilov:2019}.

\section{Berezinskii pairing in Dirac semimetals}
\label{sec:Ber-Dirac}

The other example of the Berezinskii state is related to the pairing in Dirac semimetals~\cite{Wehling-Balatsky:rev-2014,Armitage-Vishwanath:2017-Rev}.
As was shown in ref.~\cite{Sukhachov-Balatsky:2019-Dirac}, both even- and odd- frequency pairing states are possible for interacting Dirac quasiparticles. However, one would need to reach the critical coupling strength $g> g_{crit}$ to develop superconducting order in Dirac materials in the case of chemical potential being exactly at the Dirac point. For attractive interaction, the preferred pairing channel is the s-wave, spin-singlet, intra-chirality, even-frequency state. The odd-frequency pairing is possible for a repulsive frequency-dependent interaction.
In the case of s-wave, spin-singlet, and inter-chirality pairing, the matrix structure of superconducting gap is $\hat{\Delta}_{\rm odd}(\omega_n) = \gamma^0\gamma_5 \Delta(\omega_n)$, where $\gamma_5=i\gamma_0\gamma_x\gamma_y\gamma_z$ as well as $\gamma^0$ and $\bm{\gamma}$ are the Dirac matrices.
The corresponding projected anomalous propagator at the charge neutrality point is
\begin{eqnarray}
\label{F-odd-Dirac}
&&F^{o}(\omega_n, \mathbf{k})=-\Delta(\omega_n) \nonumber\\
&& \times \frac{\omega_n^2 -\epsilon_{\mathbf{k}}^2 +\left|\Delta(\omega_n)\right|^2}{\left(\omega_n^2+\epsilon_{\mathbf{k}}^2\right)^2 +2\left|\Delta(\omega_n)\right|^2\left(\omega_n^2-\epsilon_{\mathbf{k}}^2\right) +\left|\Delta(\omega_n)\right|^4},
\end{eqnarray}
where $\epsilon_{\mathbf{k}}=v_Fk$ stands for the quasiparticle dispersion relation in Dirac semimetals, $\mathbf{k}$ is momentum, and $v_F$ is the Fermi velocity.
In the vicinity of the critical phase transition $T\approx T_c$,
\begin{eqnarray}
\label{F-odd-Dirac-Tc}
F^{o}(\omega_n, \mathbf{k})\approx -\Delta(\omega_n) \frac{\omega_n^2 -\epsilon_{\mathbf{k}}^2}{\left(\omega_n^2 +\epsilon_{\mathbf{k}}^2\right)^2}.
\end{eqnarray}
Then, as an example, a simple odd-frequency ansatz $\Delta(\omega_n) = \Delta_0 \omega_n$ yields
\begin{eqnarray}
\label{B-odd-Dirac-Tc}
B_{\mathbf{k}}=2\sum_{n=0}^{\infty}\frac{1}{\omega_n} F^{o}(\omega_n, \mathbf{k}) = -\frac{\Delta_0}{4\cosh^2{(\epsilon_{\mathbf{k}}/2)}}.
\end{eqnarray}

\begin{figure}
  \includegraphics[width=\columnwidth]{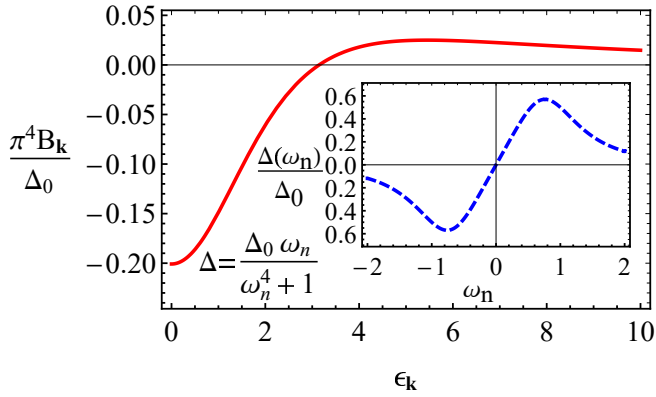}%
  \caption{\label{fig:Odd-dirac-beta}\col
    The dependence of the order parameter $B_{\mathbf{k}}$ on $\epsilon_{\mathbf{k}}$. The inset shows $\Delta(\omega_n) = \alpha \omega_n/\left(\omega_n^4+1\right)$ as a function of $\omega_n$. Note that $\epsilon_{\mathbf{k}}$ is measured in units of temperature, which was set to unity.}
\end{figure}

The results for a more complicated ansatz $\Delta(\omega_n) = \Delta_0 \omega_n/\left(\omega_n^4+1\right)$ are shown in Fig.~\ref{fig:Odd-dirac-beta}. Note that, unlike $B_{\mathbf{k}}$ given in Eq.~(\ref{B-odd-Dirac-Tc}), the correlation function $B_{\mathbf{k}}$ in  Fig.~\ref{fig:Odd-dirac-beta} changes sign.

\section{Conclusion}
\label{sec:conclusions}

In this paper, we introduced the concept of the time-independent correlators for even- and odd-frequency pairing states, which in the case of fermions read as $A_{\alpha \beta}$ and $B_{\alpha \beta}$, respectively.
In the case of bosons, the corresponding correlators $\mathcal{A}_{\alpha \beta}$ and $\mathcal{B}_{\alpha \beta}$ are defined in a similar way albeit should include a derivative over temperature.
These correlators provide an alternative approach for capturing the existence of the odd-frequency Berezinskii pairing channel and can be used as the order parameters for this hidden order.

To illustrate the concept of the time-independent correlators, we considered $B_{\alpha\beta}$ in the case of Majorana fermions and Dirac semimetals. It is found that the time-independent correlators are able to effectively capture the dependence on the energy scale. For example, in the case of Majorana fermions, $B_{\alpha\beta}$ is constant for noninteracting Majorana modes and scales as $1/J$ for large intra-wire interaction scales $J$. On the other hand, $B\sim 1/J^{1/2}$ for interacting Majorana modes in the SYK model.
The time-independent correlator $B_{\alpha\beta}$ can be also introduced for the Berezinskii pairing in Dirac semimetals. In particular, the latter state could be realized for a repulsive interaction in the inter-chirality, s-wave, spin-singlet channel. Interaction potential or, equivalently, the profile of the gap function strongly affects the dependence of the time-independent correlator on momentum.

We believe that our results will provide new tools for investigating hidden quantum time orders, allow constructing order parameters for the odd-frequency Berezinskii pairing, and might help to identify it in experiments.

\section{Acknowledgements}
We are grateful to E. Langman for useful discussions. Work is supported by VILLUM FONDEN via the Centre of Excellence for Dirac Materials (Grant No.  11744), Knut and Alice Wallenberg Foundation (Grant No. KAW 2018.0104), and the European Research Council ERC-2018-SyG HERO.

\end{document}